# Electronic structure*s* of air-exposed few-layer black phosphorus by optical spectroscopy

Fanjie Wang, Guowei Zhang, Shenyang Huang, Chaoyu Song, Chong Wang, Qiaoxia Xing , Yuchen Lei, and Hugen Yan[*]

Department of Physics, State Key Laboratory of Surface Physics and Key Laboratory of Micro and Nano Photonic Structures (Ministry of Education), Fudan University, Shanghai 200433,China

The electronic structure of few-layer (FL) black phosphorus (BP) sensitively depends on the sample thickness, strain and doping. In this paper, we show that it's also vulnerable to air-exposure. The oxidation of BP caused by air-exposure gives several optical signatures, including the broadening of resonance peaks and increased Stokes shift between infrared (IR) absorption and photoluminescence (PL) peaks. More importantly, air exposure causes blue shifts of all resonance peaks in IR absorption and PL spectra, with more prominent effects for thinner samples and higher order subband transitions. Our study alludes a convenient and exotic way for band-structure engineering of FL-BP through controllable air-exposure or defect creation.

## I. INTRODUCTION

Black phosphorus (BP) has been demonstrated as a promising candidate for electronic and photonic devices [1-3], benefiting from its unique physical properties, such as in-plane anisotropy [4, 5], widely tunable direct bandgap from mid-infrared to visible frequency range [6-8] and relatively high carrier mobility [9, 10]. Unfortunately, the unpaired electron on the surface makes it reactive to air, causing degradation in samples. Although the mechanism of degradation is still inconclusive, a growing consensus shows photo-oxidation, aided by moisture, is the main cause [11-15]. In particular, ultraviolet photons play a major role [16]. Oxygen breaks the P-P bonds and irreversibly converts BP into $P_xO_y$ compounds. Moisture accelerates the subsequent decomposition of $P_xO_y$. BP's instability in air hampers the characterization of its intrinsic properties and its device performance. Therefore, it is essential to interrogate the physical consequences of air-exposure.

The respective influence of light, oxygen, humidity on the degradation of few-layer (FL) black phosphorus (BP) has already been addressed in recent studies [17-26]. Corresponding impacts have also been studied by atomic force microscopy (AFM), X-ray photoelectron spectroscopy (XPS) and Raman spectroscopy, showing that many visible bubbles appear on the surface after a certain time of exposure to air, with remarkably reduced intensity ratio of $A_g^1/A_g^2$ Raman modes [27]. However, AFM can only be used for the surface topography identification, which is not suitable for the early stage diagnostics. In addition, caution must be taken when characterizing the degradation by Raman spectroscopy. Due to many independent variables such as lattice orientation, thickness and excitation wavelengths, Raman response shows complicated behaviors and contradictions still exist [28-31]. Though many efforts have been devoted to exploring the mechanisms and products of degradation, little is known regarding the band structure of air-exposed BP.

---

[*] hgyan@fudan.edu.cn

In this paper, we systematically study the degradation-induced changes in the band structure of mechanically exfoliated FL-BP, using infrared (IR) absorption and photoluminescence (PL) spectroscopy. New insights are gained for the degradation process and its consequences. From the measured IR absorption and PL spectra, we can directly trace the evolution of the (optical) bandgap and higher order subband transitions. Interestingly, the degradation induces blueshift of all optical resonances, with shift rates subband- and layer-dependent. Clear signs of blueshift show up within a few minutes of air-exposure, underlining the rapidity of degradation and the sensitivity of our characterization technique. Our results are expected to provide a reliable and sensitive determination of the band structure in air-exposed BP and pave the way for quick assessment of device quality by optical means. Meanwhile, our study demonstrates a novel scheme of band structure engineering through controllable air-exposure.

## II. EXPERIMENTAL METHODS

FL-BP samples were mechanically exfoliated on a transparent polydimethylsiloxane (PDMS) substrate in a glove box. The layer (L) number was first estimated by optical contrast and further verified by IR absorption spectroscopy. Samples were protected in an enclosed Linkam chamber (Linksy82) with continuously purged nitrogen during the IR absorption and PL measurements.

As an anisotropic material, BP has two characteristic crystal orientations: armchair (AC) and zigzag (ZZ), as illustrated in Fig. 1a, corresponding to the strongest and weakest optical response, respectively. The strong polarization- and layer-dependent absorption of BP allows for the fast and accurate determination of crystal orientation and layer number by IR absorption spectroscopy [32, 33]. In this work, all of the IR absorption spectra were obtained with incident light polarized along the AC direction. Besides, PL spectroscopy was also used for sample characterization with thickness $\leqslant$ 3L.

All IR absorption spectra of FL-BP samples were recorded using a Bruker FTIR spectrometer (Vertex 70v) equipped with a Hyperion 2000 microscope. A tungsten halogen lamp was used as the light source to cover the broad spectral range of 3750–11000 $cm^{-1}$ (0.36–1.36 eV), in combination with a liquid nitrogen cooled MCT detector. The lower bound cutoff photon energy is restricted by the PDMS substrate. The incident light was focused on BP samples using a 15X IR objective, with the polarization controlled by a broadband ZnSe grid polarizer.

The PL measurements on FL-BP samples were conducted using an Andor AR500i spectrometer equipped with the InGaAs detector, in conjunction with a Nikon microscope and a 50X objective. A 532 nm laser was used for excitation. Samples were kept under the same low laser excitation power to avoid laser-induced heating and damage during measurements.

## III. RESULTS AND DISCUSSIONS

### A. Optical characterization of FL-BP

Owing to the layer-layer interactions and quantum confinement in the out-of-plane direction, conduction and valence bands of the *N*-layer BP split into *N* 2-dimensional (2D) subbands. As an example, Fig. 1b illustrates the simplified band structures of a 2L BP, where $c_1$, $c_2$ and $v_1$, $v_2$ denote the quantized conduction and valence subbands and the vertical arrows illustrate the optical transitions between $c_1$ and $v_1$, and $c_2$ and $v_2$, respectively. Here we define $E_{nn}$ as the nth order subband transition (transition from $v_n$ to $c_n$), since interband transitions in symmetric quantum wells with normal light incidence are only allowed for the same subband index $n$.

Figure 1d shows the IR extinction (1-T/$T_0$) spectra of a freshly exfoliated 3L BP (the optical image is shown in Fig. 1c) on a PDMS substrate, where $T_0$ and T denote the light transmission through the PDMS without and with BP on it, respectively. Obviously, the IR absorption is dominated by significant exciton effect of the $E_{11}$ transition, as expected in low-dimensional semiconductors, while $E_{22}$ and $E_{33}$ transitions are beyond our measurement range. Moreover, strong PL emission (the inset in Fig. 1d) was observed in this sample, with its peak energy (0.85 eV) almost the same as the absorption peak, showing a negligible Stokes shift (see Fig. 4a). This further confirms the high quality of pristine FL-BP samples.

### B. Subband- and thickness-dependent blueshift

To characterize the impacts of air-exposure on the electronic structure of FL-BP, we monitored the evolution of IR and PL spectra of FL-BP samples in air over time. All samples were deliberately exposed to air for a certain amount of time, and subsequently protected in a chamber by purging nitrogen gas for optical measurements. We checked the effectiveness of the protection method, the IR absorption spectrum barely changes over time in nitrogen gas environment, in striking contrast to air-exposed samples. This confirms the effectiveness of $N_2$ gas protection and controllable air-exposure is feasible. See the Supplemental Material for the efficient nitrogen protection [34].

We first focus on the early stage of the air exposure, in which the BP samples don't visually show any inhomogeneity from a microscope optical image. In the later stage, as discussed in the final part of the paper, the sample thickness will be reduced inhomogeneously. Figures 2a and 2b show the IR absorption spectra of 3L and 8L BP in air at different exposure times. Both BP samples were on the same PDMS substrate and the same exposure condition was ensured. For clarity, only three spectra (0 min, 5 min, 20 min) are shown here. The IR absorption spectra of fresh BP samples (black curves) exhibit sharp and narrow exciton resonances, associated with optical transitions between quantized valence and conduction subbands. The results indicate that FL-BP samples are very sensitive to air, only a few minutes of exposure cause apparent modification of the sample properties: the remarkable reduced peak height and broadened linewitdth (red and blue curves in Figs. 2a and 2b) are directly attributed to the introduction of defects [35]. More interestingly, the degradation causes apparent blueshift of peaks. Take the 3L sample for example, the resonance peak gradually shifts from 0.87 eV to 1 eV (Fig. 2a). The scenario reminds us of possible band structure tuning for FL-BP through defect engineering. The IR spectra are clear indicators of the sample quality. In the contrary, optical microscope images barely show any discernable change at this stage.

Figure 2c shows peak energies for $E_{11}$ and $E_{22}$ transitions of the 8L BP as a function of exposure time, whose typical spectra are displayed in Fig. 2b. We can see that $E_{22}$ shifts much faster than $E_{11}$. Figure 2d compares the $E_{11}$ transitions for the 3L and 8L sample shown in Figs. 2a and 2b, and the 3L sample exhibits more drastic shift. The peak shift can be fairly compared because both 3L and 8L samples are on the very same substrate and the exposure conditions are always the same. We performed similar studies for many samples with different thickness (See the Supplemental Material for more examples [34]). All measurements show the same qualitative behaviors, i.e., air exposure causes blueshift of all resonance peaks and higher order subband transitions and thinner samples are more sensitive [36].

The band structure of pristine FL-BP can be approximated as that of an infinite square quantum well. Due to the relatively strong layer-layer interaction, the effective mass of carriers in the perpendicular direction is relatively small. Therefore, the band gap and higher order optical transitions are sensitive to the sample thickness (quantum well width). By the same token, the band structure will also be sensitive to the modification of quantum well profile induced by the oxidation of the surface layer. During the early stage of the degradation, it's reasonable to assume that only the top layer is partially oxidized and other layers underneath are intact. Therefore, a potential barrier for carriers forms between the top layer and the rest of the sample. The inset in Fig. 3a shows the energy diagram of pristine BP and its oxide. Pristine BP acts as a perfect infinite square quantum well. The appearance of oxide modifies the perfect structure and introduces a barrier with height $\Delta$, as shown in Fig. 3a. This certainly modifies the electronic structure and tunes the optical transitions.

We simulate the blueshift of the optical transitions numerically. We assume the $N$-1 layers underneath are intact, and only the top layer oxidized homogeneously. The severity of the oxidation is characterized by the barrier height $\Delta$: the larger the barrier, the more severe the oxidation is. We extracted the ground and excited states in this modified quantum well by numerically solving the one-dimensional Schrodinger equation under the effective-mass approximation [4, 7] (See the Supplemental Material for details [34]). Figures 3b and 3c summarize the transition energy versus oxide barrier height, where we assumed the thickness of the oxide layer equals single layer BP thickness without loss of generality. As expected, the results are consistent with our experimental findings. First of all, all peaks shift to higher energy, since the potential barrier introduced by degradation effectively enhances the 2D quantum confinement, or equivalently, reduces the effective thickness of the quantum well. Secondly, the higher order subband envelope wave-function (in the form of $sin(n\pi z/L)$ for a perfect infinite square well, where $n$ is the subband order, $L$ is the sample thickness) has larger amplitude in the vicinity of the degraded surface, hence the energy eigenvalues of higher order subbands shift more than that of lower order. This explains the fact that $E_{22}$ shifts faster than $E_{11}$ (Fig. 3b) for the same sample. Thirdly, by assuming the same degradation rate of the top layer for different thickness samples, the thinner samples are certainly more sensitive to degradation and the peaks show higher shift rates than that for thicker samples (Fig. 3c). The simulated results are in very good qualitative agreement with our experimental findings. It should be admitted though, in this model, for simplicity, we assume all few-layer samples have the same degradation rate of the surface layer and this can already explain the faster shift of thinner samples. Of course, this hypothesis deserves caution, since theoretical studies have predicted that thinner samples (particularly, 1L or 2L) tend to react faster due to a better match between the band gap and oxide energy levels [35, 37]. Since current study doesn't focus on the thinnest samples (1L and 2L), our hypothesis is still valid to a large extent.

In spite of more defects, we can in principle take advantage of the controllable degradation for band structure engineering. The band gap of pristine FL-BP can not vary continuously when the layer number changes. However, in conjunction with controllable air-exposure of the samples, additional continuous tuning of the band gap can be achieved and optical resonance is likely to cover a wide spectral range continuously with both pristine and slightly degraded FL-BP samples. The tuning of band structure through oxidation can be applied to other 2D materials as well, as recently demonstrated in FL InSe [38]. As long as the effective mass of carriers in the *z*-direction is relatively small, the tuning will be efficient. In this regard, FL-BP is a model system to demonstrate such tuning.

### C. Stokes shift

In addition to blueshift of resonance peaks, the enhanced IR absorption and PL peak energy difference was also observed during our measurements. This energy difference, termed as Stokes shift, is quite common in semiconductor optics [39]. Figure 4a shows a direct comparison of absorption and emission spectra for the same 3L sample, each set of IR absorption and PL spectra was obtained under the same sample quality. The pristine sample showed narrow IR absorption and PL peaks, and their peak energies match well. Both of IR absorption peak and PL peak shift to higher energies after air-exposure, as shown in Fig. 4b. More importantly, a large redshift of PL peak energy relative to IR absorption peak appeared after it was exposed to air, and the energy difference increased with increasing exposure time (see Fig. 4c). It is associated with defects or impurity states below the bandgap. Luminescence process favors the lowest energy for recombination, and defect or impurity states below the bandgap tend to attract the PL peak to the lower energy side of the absorption peak. Therefore, Stokes shift is an indicator of sample quality. The enhancement of the Stokes shift of our BP samples during air-exposure is fully consistent with sample degradation.

### D. Inhomogeneous thinning

After we discuss the early stage of the degradation, now we can take a closer look at the later stage. Typically, the sample starts to visually show inhomogeneity under an optical microscope. Figure 5a shows a freshly cleaved 3L-BP sample, the layer number was verified by IR absorption spectrum (inset of Fig. 5a). After one-hour exposure in air, areas with different optical contrast emerged (Fig. 5b), which indicates the thickness of the sample was reduced in certain areas. Indeed, PL spectra confirm this assertion. We collected PL spectra at locations marked by black triangle (P1) and red star (P2) which refer to two different layer numbers according to the optical contrast. As shown in Fig. 5c, the PL spectrum at P2 (red curve) has a peak at 1.15 eV which originates from a 2L BP. The slight blueshift from a pristine 2L BP (dashed line) might result from the early stage oxidation. The PL peak at P1 (black curve), which is from the 3L part, is nearly quenched owing to a long time exposure in air. These results demonstrate that the 3L sample partly turned into 2L, with the rest of 3L strongly degraded and the newly emerged 2L still of relatively high quality. The flake therefore constitutes two types of thickness simultaneously. Of course, this process keeps going and the thickness of the sample can be further reduced inhomogeneously. Eventually, the whole piece of BP flake is decomposed, possibly leaving some faint vestiges. The spectrum of an original 5L sample with coexistence of 3, 4 and 5L BP after degradation is shown in Supplement Material [34]. The coexistence of multiple peaks in the IR spectrum indicates the inhomogeneous degradation and sample thickness reduction. Now a full picture of air-exposure for FL-BP emerges: Initially, the top layer of a *N*-layer BP is slightly oxidized and a potential barrier forms between the top layer and the rest, which shifts the resonances to higher energy; with the reaction progressing, some parts of the oxidized top layer may be decomposed by moisture, leaving a relatively fresh *N*-1 layer coexisting with the degraded *N*-layer; oxidation progresses further for the fresh *N*-1 layer and eventually the whole sample is decomposed.

## IV. CONCLUSION

In summary, we demonstrated that air-exposure has a strong effect on the electronic structure of BP. The band gap and higher order subband transitions shift to higher energies. A quantum well model can well account for our observations. The signatures of early-stage degradation include blueshift and broadening of resonance peaks, as well as Stokes shift in PL. Later stage degradation causes layer number reduction, typically inhomogeneously. Our study not only provides a sensitive means to characterize the quality of BP samples, but also paves a way for band structure tuning through controllable defect engineering.

## ACKNOWLEDGMENTS

H.Y. is grateful to the financial support from the National Young 1000 Talents Plan, National Natural Science Foundation of China (grants: 11874009, 11734007), the National Key Research and Development Program of China (grants: 2016YFA0203900 and 2017YFA0303504), Strategic Priority Research Program of Chinese Academy of Sciences (XDB30000000), and the Oriental Scholar Program from Shanghai Municipal Education Commission. Part of the experimental work was carried out in Fudan Nanofabrication Lab. G.Z. acknowledges the financial support from the National Natural Science Foundation of China (grant: 11804398) and Open Research Fund of State Key Laboratory of Surface Physics. C.W. is supported by the National Natural Science Foundation of China (Grant: 11704075).

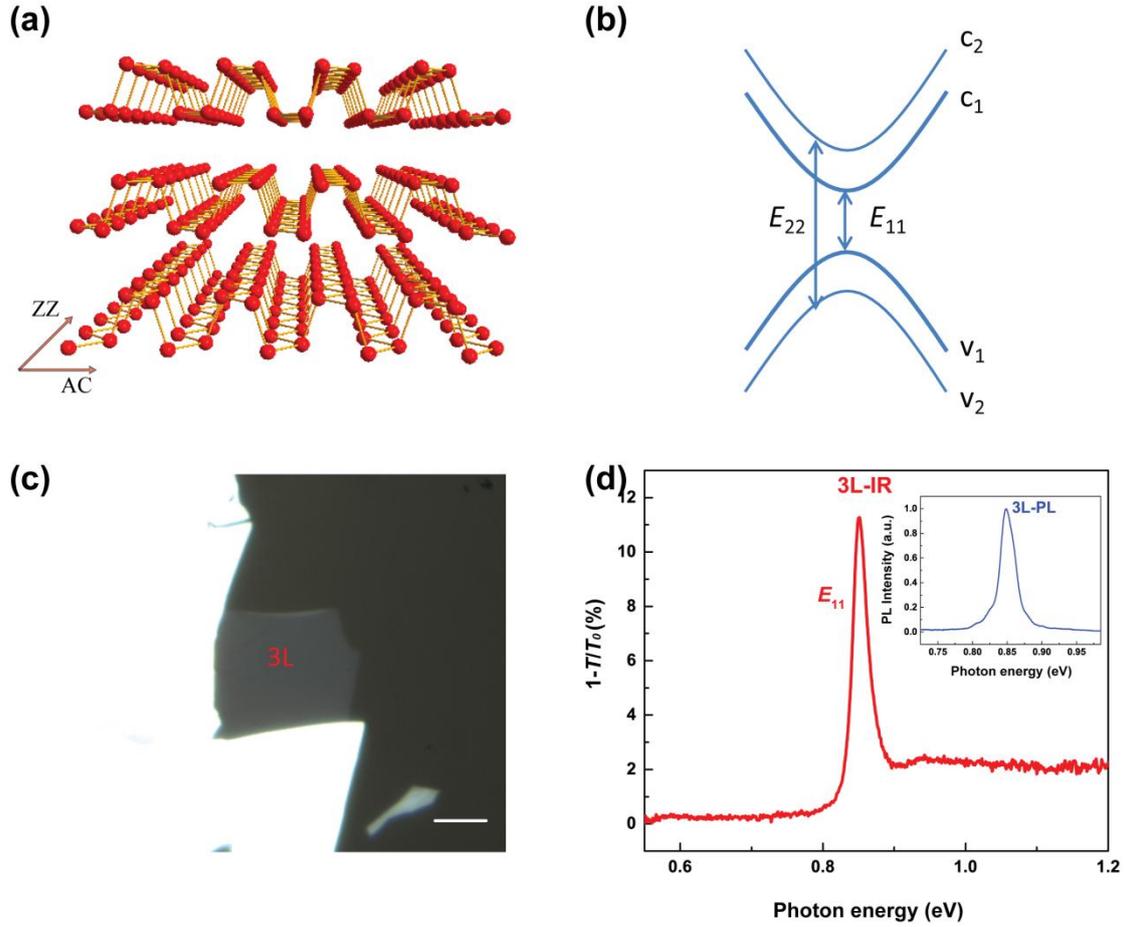

FIG. 1. Optical characterization of pristine BP samples. (a) Crystal structure of a 3L BP. (b) The qualitative energy levels of the 2L BP, vertical arrows $E_{11}$ and $E_{22}$ denote transitions $v_1 \rightarrow c_1$ and $v_2 \rightarrow c_2$, respectively. (c) Optical image of a 3L BP exfoliated on PDMS. Scale bar, 10μm. (d) IR absorption spectrum of the 3L BP. $T_0$ and T denote the light transmission of the bare PDMS and the sample on the PDMS, respectively. Inset: PL spectrum of the same 3L BP.

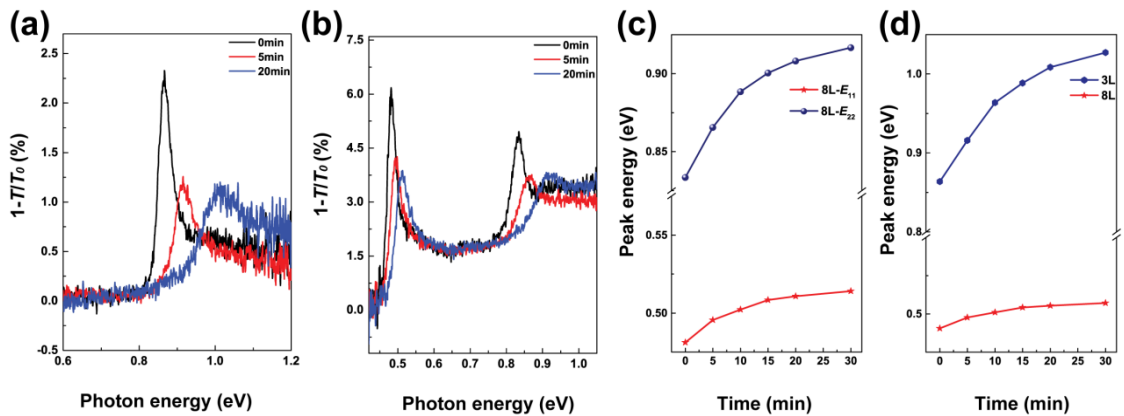

FIG. 2. Subband- and layer-dependent blueshift of (a) 3L and (b) 8L BP. (c) $E_{11}$ and $E_{22}$ peak energies of the 8L sample as a function of exposure time; for the same sample, $E_{22}$ peak moves faster than $E_{11}$. (d) $E_{11}$ peak energies of the 3L and 8L sample extracted from Figs. 2a and 2b; the 3L sample peak moves faster.

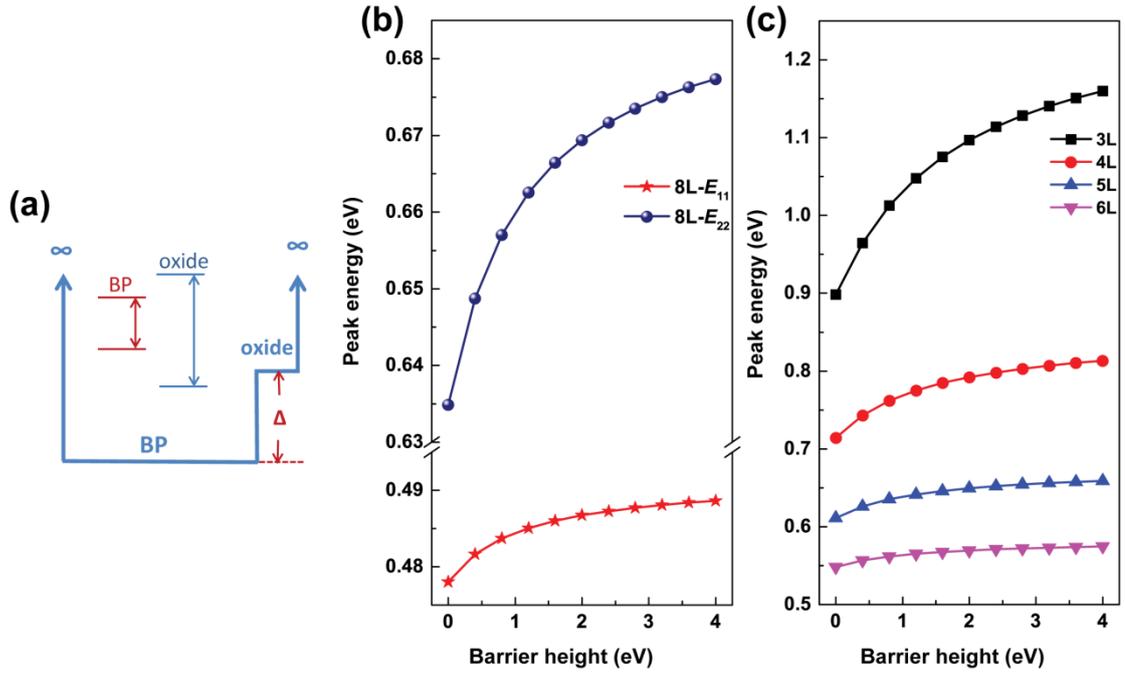

FIG. 3. The mechanism of blueshift. (a) The adjusted quantum well. Inset: The energy diagram of BP and its oxide. Δ denotes the effective barrier height. (b) Calculated $E_{11}$ and $E_{22}$ transition energies of 8L BP as a function of barrier height. (c) Calculated $E_{11}$ peak energies versus oxide's barrier height for BP with various layer numbers.

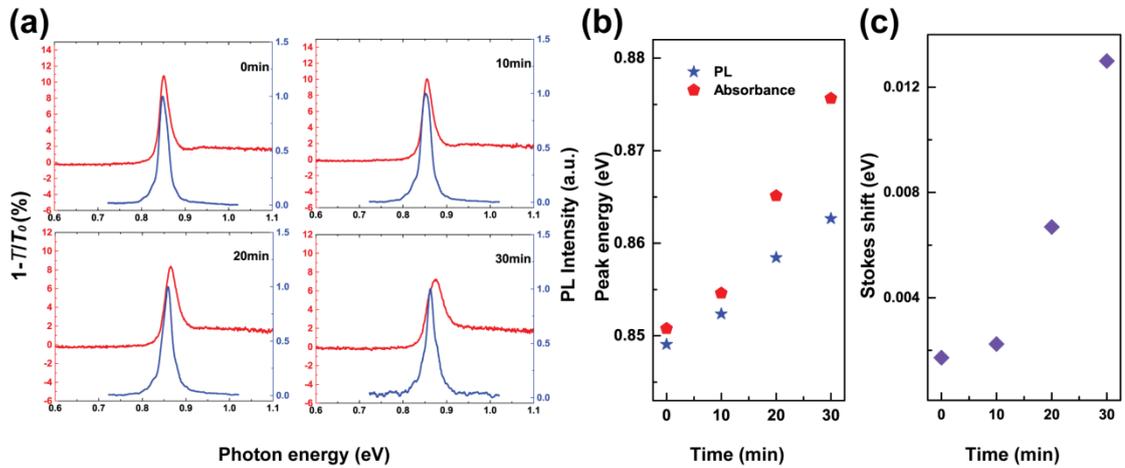

FIG. 4. Defect-induced Stokes shift. (a) Comparison of IR absorption and PL spectra of a 3L BP on the PDMS substrate. The spectra are shifted vertically for clarity. (b) Peak energies extract from Fig. 4a. (c) Stokes shift during the degradation, indicating the accumulation of defects over time.

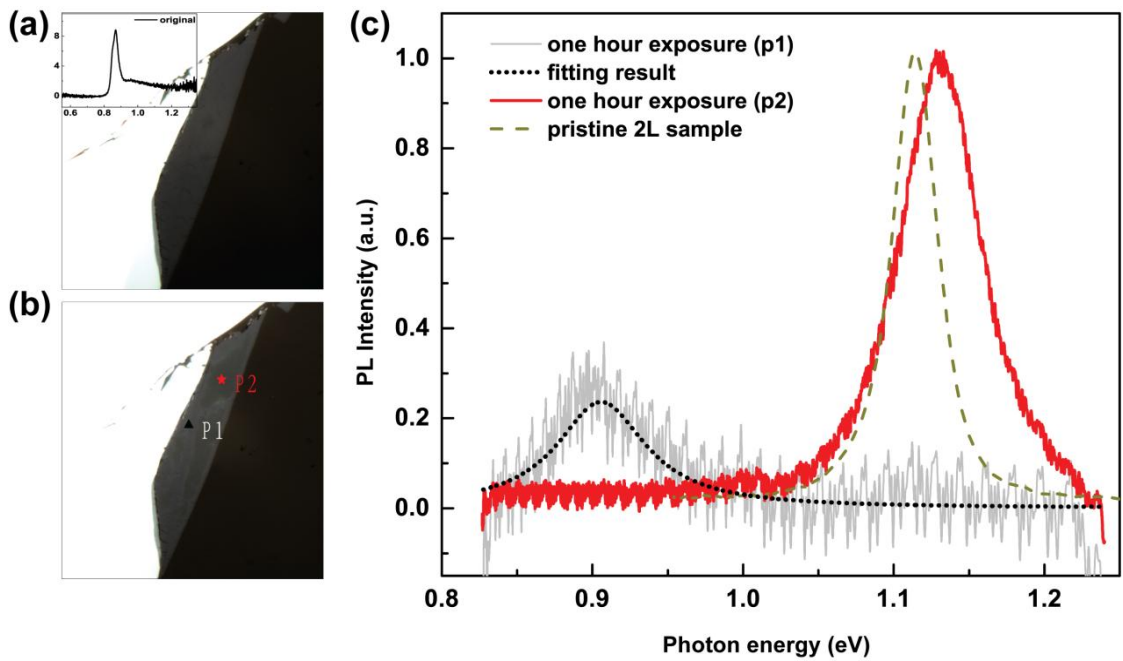

FIG. 5. Inhomogeneous layer thinning. (a) Optical microscope image of an original cleaved 3L BP sample on the PDMS substrate. Inset shows the IR absorption spectrum. (b) Optical microscope image of Fig. 5a after one hour exposure to air: P1 represents the location marked by black triangle and P2 represents the location marked by red star. (c) PL spectra of typical locations marked in (b). Two different peaks refer to 3L and 2L BP, respectively. A pristine 2L PL spectrum is shown for comparison.